\documentclass[preprint,superscriptaddress,showpacs,preprintnumbers,amsmath,amssymb]{revtex4-1}

\usepackage{amsmath}
\usepackage{amssymb}
\usepackage{graphicx}
\usepackage{morefloats}
\usepackage[usenames,dvipsnames]{xcolor}
\usepackage{blkarray}
\usepackage{verbatim}
\usepackage{hyperref}
\usepackage{color}

\begin{document}

\title{The Role of Noise in the Spatial Public Goods Game}

\author{Marco Alberto Javarone}
\email{marcojavarone@gmail.com}
\affiliation{Department of Mathematics and Computer Science, University of Cagliari, Cagliari (Italy)}
\author{Federico Battiston}
\affiliation{School of Mathematical Sciences, Queen Mary University of London, Mile End Road, E1 4NS, London (UK)}

\date{\today}

\begin{abstract}
In this work we aim to analyze the role of noise in the spatial Public Goods Game, one of the most famous games in Evolutionary Game Theory. 
The dynamics of this game is affected by a number of parameters and processes, namely the topology of interactions among the agents, the synergy factor, and the strategy revision phase.
The latter is a process that allows agents to change their strategy. Notably, rational agents tend to imitate richer neighbors, in order to increase the probability to maximize their payoff. By implementing a stochastic revision process, it is possible to control the level of noise in the system, so that even irrational updates may occur.
In particular, in this work we study the effect of noise on the macroscopic behavior of a finite structured population playing the Public Goods Game. We consider both the case of a homogeneous population, where the noise in the system is controlled by tuning a parameter representing the level of stochasticity in the strategy revision phase, and a heterogeneous population composed of a variable proportion of rational and irrational agents.
In both cases numerical investigations show that the Public Goods Game has a very rich behavior which strongly depends on the amount of noise in the system and on the value of the synergy factor.
To conclude, our study sheds a new light on the relations between the microscopic dynamics of the Public Goods Game and its macroscopic behavior, strengthening the link between the field of Evolutionary Game Theory and statistical physics.
\end{abstract}

\maketitle
\section{Introduction}
Nowadays, Evolutionary Game Theory (hereinafter EGT)~\cite{szolnoki01,tomassini01,perc01,nowak01,nowak02,moreno01,masuda01} represents an emerging field, whose interdisciplinarity makes it of interest for scientists coming from different communities, from biology to social sciences.
The emergence of cooperation~\cite{nowak04} in structured populations~\cite{nowak03}, in particular in games characterized by a Nash equilibrium of defection, is one of the main topics studied in this field. For instance, different mechanisms and strategies have been proposed to foster cooperation~\cite{meloni01,antonioni01,perc06} in paradigmatic games representing pairwise interactions, such as the Prisoner's Dilemma~\cite{perc05}, and group interactions, namely the Public Goods Game (hereinafter PGG)~\cite{perc03}.
Since these models are mainly studied by means of numerical simulations, they often lack of an analytical description.
However, the latter can be provided once some specific assumptions are introduced, as for instance considering memory-aware agents (i.e. able to save their payoff over time)~\cite{javarone01}.
In the modern area of complex systems~\cite{barabasi01,szabo03}, investigations driven by statistical physics~\cite{huang01}, trying to relate the macroscopic emergent behavior to the microscopic dynamics of the agents, are useful to get insights on a wide range of topics, from socio-economic systems to biological phenomena.
Therefore, in this work we aim to provide a description of the PGG by the lens of statistical physics, focusing in particular on the impact of noise in the population dynamics. Notably, the noise is controlled by a parameter adopted in the strategy revision phase (SRP), i.e., the process that allows agents to revise their strategy. The SRP can be implemented in several ways, e.g. considering rational agents that aim to increase their payoff. Usually, rational agents tend to imitate their richer neighbors, while irrational agents are those that randomly change their strategy. Remarkably, the level of noise in the system strongly affects the macroscopic behavior of a population.
Although previous works (e.g.~\cite{szabo01,szabo02,szabo04,moreno02}) focused on this topic (i.e. the role of noise) in this game, a complete analysis is still missing. In particular, we study the effect of noise in two different scenarios. We first consider the case of a homogeneous population, where the intensity of noise in the system is controlled by tuning the level of stochasticity of all agents during the SRP, by means of a global parameter. The latter is usually indicated by $K$, and defined as temperature or as an inverse degree of rationality~\cite{perc03}.
Then, we consider a heterogeneous population, characterized by the two species: rational and irrational agents. While the former take their decision considering the payoff of their neighbors, the latter take decisions randomly.
In both cases, we study the macroscopic dynamics of the population and the related equilibria, achieved for different amount of noise and values of the synergy factor. It is worth to emphasize that the synergy factor, before mentioned, is a parameter of absolute relevance in the PGG, as it supports cooperative behaviors by enhancing the value of the total contributions provided by cooperators.
Eventually, we recall that the influence of rationality in the PGG has been studied, by a probabilist approach, in~\cite{miguel01} where authors implemented agents able to select (with a given probability) between a rational and an irrational behavior.
The remainder of the paper is organized as follows: Section~\ref{sec:model} introduces the dynamics of the PGG, and the setup of our model. Section~\ref{sec:results} shows results of numerical simulations. Eventually, Section~\ref{sec:conclusions} ends the paper.
\section{The Public Goods Game}\label{sec:model}
The PGG is a simple game involving $N$ agents that can adopt one of the following strategies: cooperation and defection. Those playing as cooperators contribute with a coin $c$ (representing the individual effort in favor of the collectivity) to a common pool, while those playing as defectors do not contribute. Then, the amount of coins in the pool is enhanced by a synergy factor $r$, and eventually equally divided among all agents. 
The cooperators' payoff (i.e. $\pi^{c}$) and that of defectors (i.e. $\pi^{d}$) read
\begin{equation}\label{eq:pgg_payoff}
\begin{cases}
\pi^{c} = r \frac{N^c}{G} - c\\
\pi^{d} = r \frac{N^c}{G}
\end{cases}
\end{equation}
\noindent where $N^c$ is the number of cooperators among the $G$ agents involved in the game, $r$ synergy factor, and $c$ agents' contribution. Without loss of generality, we set $c = 1$ for all agents. 
Let us proceed focusing the attention on the synergy factor $r$~\cite{perc03}. In well-mixed populations of infinite size, where agents play in group of $G=5$ players, the two absorbing states appear separated at a critical point $r_{\rm{wm}}=G=5$. Notably, these populations fall into full defection for $r<r_{\rm{wm}}$ and into full cooperation for $r>r_{\rm{wm}}$. 
When agents are placed in the nodes of a network, surprisingly, some cooperators can survive for values of $r$ lower than $r_{\rm{wm}}$. This effect is known as \textit{network reciprocity}~\cite{nowak03,moreno02,perc04}, since a cooperative behavior emerges as a result of the same mutualistic interactions taking place repeatedly over time. At the same time, the network structure allows a limited number of defectors to survive also beyond $r=r_{\rm{wm}}$.
We refer to the two critical values of $r$ at which cooperators first appear, and defectors eventually disappear from the population, respectively as $r_{c1}$ and $r_{c2}$. 
In a networked population, depending on the values of $r$ and on how agents are allowed to update their strategy, it is possible to observe different phases: two ordered equilibrium absorbing phases, where only one strategy survives (either cooperation or defection), and an active but macroscopically stable disordered phase~\cite{javarone01} corresponding to the coexistence between the two species (i.e. cooperators and defectors).
Now, it is worth to introduce the mechanism that allows the population to evolve, i.e. the process previously mentioned  and defined as strategy revision phase. In principle, the SRP can be implemented in several ways, as agents may follow different rules or may be provided with particular behaviors. 
Usually this process is payoff-driven, i.e. agents tend to imitate richer neighbors~\cite{perc03}. However, as shown in previous works social behaviors like conformity~\cite{javarone02} can have a relevant impact on the way agents choose their next strategy.
In this work we consider a payoff-driven SRP, computing the probability that one agent imitates a neighbor according to the following Fermi-like equation
\begin{equation}\label{eq:fermi_function}
W(s_y \leftarrow s_x) = \left(1 + \exp\left[\frac{\pi_y - \pi_x}{K_y}\right]\right)^{-1}
\end{equation}
\noindent where $\pi_x$ and $\pi_y$ correspond to the payoffs of two linked agents, and $s_x$ and $s_y$ indicate their strategy, i.e., that of $x$ and of $y$, respectively. Notably, equation~\ref{eq:fermi_function} is related to a SRP performed by the $y$-th agent that evaluates whether to imitate the $x$-th one.
A crucial parameter appearing in equation~\ref{eq:fermi_function} is $K_y$, which plays the role of noise and then parametrizes the uncertainty in adopting a strategy. Notably, a low noise entails agents strongly consider the difference in payoff $\Delta p = \pi_y - \pi_x$ while deciding their next strategy, while increasing the noise the payoff difference plays a more marginal role. 
In the case of a homogeneous population $K$ is equal for all individuals, so by tuning its value we are able to control the level of noise in the system.
In the limit of $K=0$, the $y$-th agent will imitate the strategy of the $x$-th agent with probability $W=1$ if $\pi_x > \pi_y$, and $W=0$ otherwise. Conversely, in the limit $K \to \infty$ the SRP becomes a coin flip, and the imitation occurs with probability $W=1/2$ no matter the value of the synergy factor. In the latter case the behavior of the PGG is analogous to that of a classical voter model~\cite{ligget01}, where imitation between a pair of selected agents takes place with probability $W=1/2$. Our study aims to confirm computationally results reported in~\cite{szabo02,moreno02}, and to evaluate the relation between $r$ and $K$, in order to provide a complete description of the PGG, from the microscopic dynamics to the global behavior of the population.

According to previous investigations, setting $K = 0.5$ is often considered as a good choice to describe a rational population, where a limited number of irrational updates may still occur. In the case of bidimensional lattices, it has been shown in~\cite{perc03} that for such $K$ the values of $r_{c1}$, at which cooperators emerge, and $r_{c2}$, where defectors completely disappears from the population, are respectively equal to $3.75$ and $5.5$.
It is also possible to consider the case of heterogenous populations where the value of $K^y$ is an agent-dependent parameter. In such scenario, the simplest way to control the noise is considering two different species of agents: those with $K^1$ and those with $K^2$.  Thus, by varying the density $f$ of one species, with $0 \le f \le 1$, it is possible to study the outcome of the model in different conditions.

Such second version is useful to analyze the behavior of a population whose agents have a different sensibility to their payoff and, from a social point of view, it allows to study the influence of rationality in driving the population towards an equilibrium (or steady state). For instance, setting $K^1=0.5$ and $K^2=\infty$, it is possible to evaluate the influence of a density $f$ of rational agents in driving the population towards a particular state.
In the sociophysics literature~\cite{loreto01,galam01}, studying the effects of social behaviors in the dynamics of a population constitutes one of the major aims. Just to cite few, investigations on nonconformity~\cite{javarone03}, extremisms~\cite{javarone04} and stubbornness~\cite{galam02} showed their relevant role in processes of opinion dynamics~\cite{loreto01,galam03,weron01,battiston01,javarone05}. Eventually, in a heterogeneous population, it would be interesting to consider more complicated cases where agents are characterized by a broad distribution of values of $K$, and can possibly change their own degree of rationality, for instance by thermalization-like processes (i.e. when two agents play, they modify their degree of rationality taking the average value of their current $K$). 

Finally, we consider an asynchronous dynamics, where the agents sequentially update their strategy. In doing so, the population dynamics can be summarized as follows:
\begin{enumerate}
\item at $t=0$ set a population with the same amount of cooperators and defectors;
\item select randomly two connected agents: $x$ and $y$; %
\item selected agents play the PGG with their communities of belonging (accumulating a payoff);
\item $y$ performs the SRP using $x$ as reference (by equation~\ref{eq:fermi_function});
\item repeat from $ii)$ until the population reaches an ordered phase (or up to a limited number of time steps elapsed).
\end{enumerate}
We indicate with $T$ the average number of updates per agents at which the dynamics has been stopped, either because the system reached an absorbing phase or because no significant macroscopic changes were observed in the system over time. The considered maximum length of the simulation is $10^6$.
Since in the PGG, the strategy of the $x$-th agent can be described by a binary variable $s_x = \pm 1$, with $+1$ representing cooperation and $-1$ defection the average magnetization~\cite{mobilia01} reads
\begin{equation}\label{eq:magnetization}
M=\frac{1}{N}\sum_{i=1}^N s_i,
\end{equation}
\noindent where $M=1$ corresponds to full cooperation, while $M=-1$ to full defection. Since we are not interested in the sign of the prevalent strategy, but only to which extent the system is ordered, we consider the absolute value of the magnetization $|M|$. From equation~\ref{eq:magnetization} it is straightforward to derive the density $\rho$ of cooperators in the population
\begin{equation}\label{eq:density_cooperators}
\rho = \frac{M+1}{2}
\end{equation}
\noindent so that we can identify the two (ordered) absorbing states corresponding to $\rho = 1$ (i.e. full cooperation) and $\rho = 0$ (i.e. full defection). In the following section, the system is analyzed by reporting the average value of $\rho$, $|M|$ and $T$ over 100 simulations.
At last, another interesting order parameter useful to detect fluctuations in the system's behavior is the standard deviation of the fraction of cooperator $\sigma(\rho)$ obtained over the different runs.
\section{Results}\label{sec:results}
We performed several numerical simulations of the PGG, for different values of the synergy factor $r$ and the noise (measured either in terms of $K$ or density of irrational agents $1-f$), in a population of $N = 10^4$ agents distributed on a bidimensional lattice with periodic boundary conditions. 

\begin{figure*}[!ht]
\centering
\includegraphics[width=0.75\textwidth]{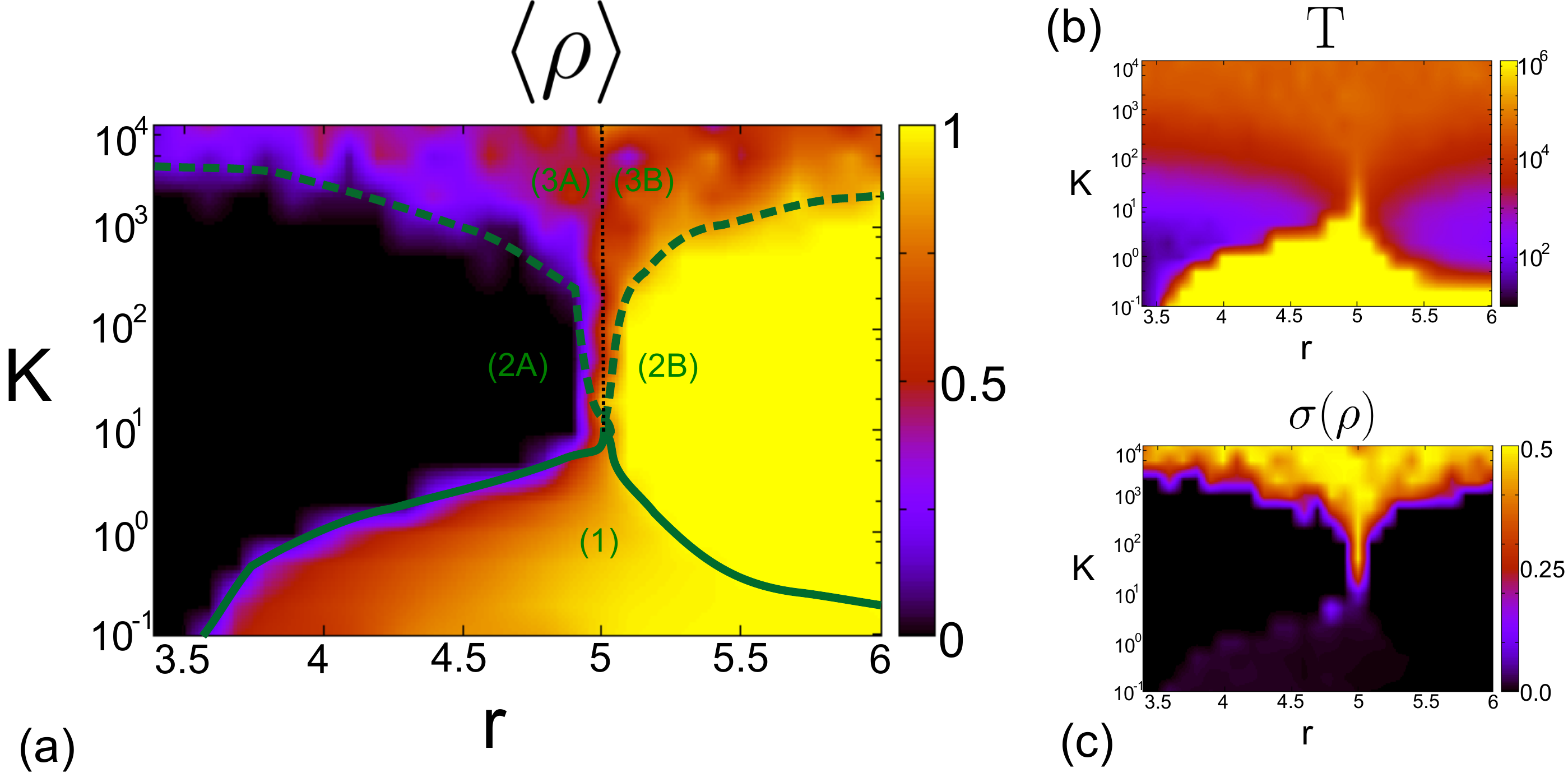}
\caption{\small (Color online). Phase diagram of the average density of cooperators $\langle \rho \rangle$ at the steady state (a), time to reach the absorbing state $T$ (b) and standard deviation of the density of cooperators $\langle \rho \rangle$ at the steady state (c) as a function of the synergy factor $r$ and the rationality $K$. Different regions are highlighted. In region (1) the system is stuck in a metastable active phase, macroscopically at the equilibrium, with coexistence of cooperators and defectors due to network reciprocity (the simulations have been stopped after $T=10^6$ updates per agent). In region (2) the system always reaches the absorbing state predicted by the well-mixed population approximation, i.e. full defection for $r<r_{\rm{wm}}=5$ and full cooperation for $r>r_{\rm{wm}}$. In region (3) both steady states become accessible with different probability, as in a biased voter model. Results are averaged over $100$ simulation runs. \label{fig:figure_1} }
\end{figure*}

\begin{figure*}[!ht]
\centering
\includegraphics[width=0.75\textwidth]{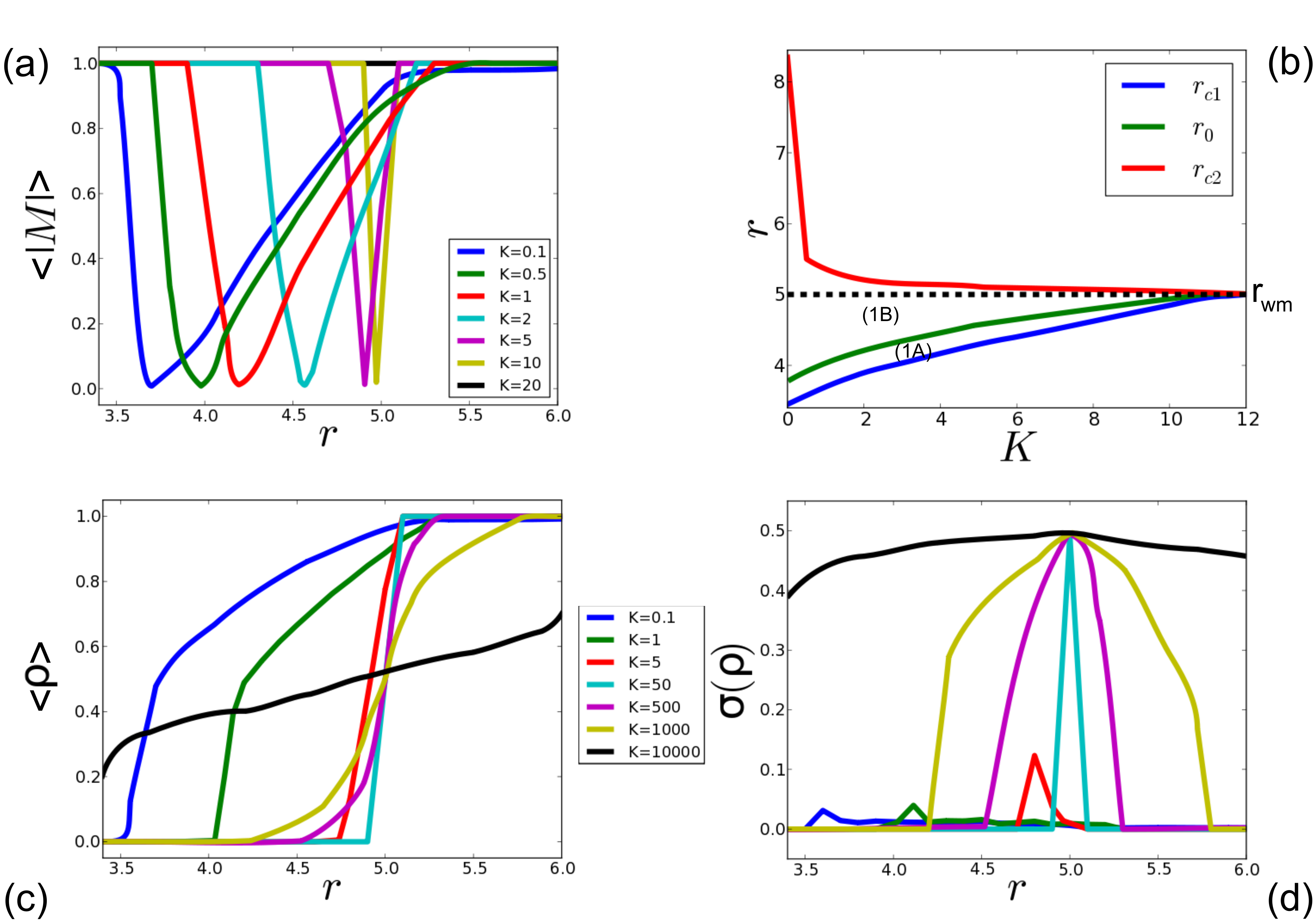}
\caption{\small (Color online) In the top panels we focus on the transition from the active phase towards the ordered phase. In (a) we show the average absolute value of the magnetization $\langle |M| \rangle$ as a function of the synergy factor $r$ for different $K$. As the temperature $K$ increases, the range of $r$ giving rise to an active phase shrinks around $r=r_{\rm{wm}}=5$ up to a critical value beyond which network reciprocity disappears. The scaling as a function of $K$ of the two extreme points of the active range, $r_{c1}$ and $r_{c2}$, as well as the value of $r_0$ for which $|M|=0$, are shown in (b). In the bottom panels we show the average density of cooperators $\langle \rho \rangle$ (c) and the standard deviation $\sigma(\rho)$ for selected values of $K$. For the three smallest temperatures the system crosses region (1), marked by a second order transition in $\langle \rho \rangle$ and small values of $\sigma$. For $K=50$, on each single run the system always reaches one of the two absorbing states. $\langle \rho \rangle$ is equal to $0$ (1) for low (high) values of $r$, but takes intermediate values around $r=5$. The transition is quite steep and $\sigma(\rho)=0$ unless around $r=5$. For higher values of $K$, for even a greater range of values of $r$ around $r=5$ both full defection or cooperation are achievable, $0<\langle \rho \rangle<1$ and $\sigma > 0$. In such regime the system behaves as a biased voter model under the external field $r-r_{\rm{wm}}$. As $K$ increases, the behavior of an unbiased voter model, no matter the value of $r$, is approached. Results are averaged over $100$ simulation runs. \label{fig:magnetization}}
\end{figure*}

\textbf{Homogeneous Populations.} Let us here show results for the homogeneous case, where the level of noise in the system is controlled by the global variable $K$ used in the SRP. We first analyze the phase diagram of the average density of cooperators $\langle \rho \rangle$ as a function of $r$ and $K$ ---Figure~\ref{fig:figure_1}.
Here, we observe that the PGG has a very rich behavior. For instance, plot \textbf{a} of Figure~\ref{fig:figure_1} shows $5$ different regions (below described) of interest when studying the density of cooperators at equilibrium.
Notably, low values of $K$ (i.e., $K < 10$) let emerge three phases as a function of $r$ in the considered range (i.e., from $3.4$ to $6.0$): two ordered phases (i.e., full defection and full cooperation) for low and high values of $r$, and a mixed phase (i.e., coexistence) for intermediate values of $r$.
Therefore, at a first glance, an order-disorder phase transition of second kind emerges crossing the region labeled $(1)$ in the first phase diagram (i.e., \textbf{a} of Figure~\ref{fig:figure_1}).
For higher values of $K$, next to $K=10$, the active phase vanishes and the population always reaches an ordered phase. A more abrupt phase transition between the two ordered phases, separating region $(2a)$ (full defection) and $(2b)$ (full cooperation), appears, resembling analytical results obtained for the well-mixed approximation, even if fluctuations are possible near the critical point $r=5$.
For greater values of $K$, the region of $r$ around $r=5$ such that both ordered states are attainable increases. In such range of values the system behaves as a biased voter model, where the absorbing states of cooperation (defection) is favored for $r>5$ ($r<5$). In the limit $K \to \infty$, the behavior of the system approaches that of a classical unbiased voter model, no matter the value of the adopted synergy factor. 
Plots \textbf{b} and \textbf{c} of Figure~\ref{fig:figure_1} confirm the main differences among the five regions of plot \textbf{a}. The former shows that the number of time steps reaches the maximum value of $T$ for intermediate $r$ entered around $r=5$ with low $K$, while $T$ is reduced when the population reaches an absorbing state (i.e., full defection or cooperation). Instead, plot \textbf{c} shows that the variance reaches a maximum value (as expected), $\sigma(\rho)=1/2$, when the PGG behaves like a voter model, while smaller non-null values are also obtained for the active phase, due to the existence of fluctuations.
In order to obtain a deeper characterization of the phase transitions occurring in the PGG, we study the average absolute value of the magnetization $|M|$, as a function of the synergy factor for different $K$ values.
As shown in plot \textbf{a} of Figure~\ref{fig:magnetization}, only for values of $K < 10$ there are values of the synergy factors $r$ such that $\langle | M | \rangle \neq 1$, since at $K \approx 10$ a more abrupt phase transition between full defection and full cooperation emerges, resembling the first-order first transition predicted analytically in the case of well-mixed population of infinite size.
Then, we note that for all $K$ values in the range $[0 \le K \le 10]$, it is possible to find a synergy factor $r$ such that $|M(r)| = 0$.
Higher values of $K$ strongly affect the PGG. Notably, the distance between the two thresholds of $r$ (i.e., $r_{c1}$ and $r_{c2}$) reduces by increasing $K$ ---see plot \textbf{b} of Figure~\ref{fig:magnetization}. It is interesting to note that these two values converge to the same point as $r = 5$ for $K \ge 10$. Furthermore, we also observe that  the value of $r_0$, for which cooperators and defectors coexist in equal number in the active phase, is always smaller than $r_{\rm{wm}}$ ---see plot \textbf{b} of Figure~\ref{fig:magnetization}.
Eventually, both plots \textbf{c} and \textbf{d} of Figure~\ref{fig:magnetization} clearly confirms the previous investigations. For instance, for $K = 10000$ the density of cooperators becomes almost flat as in a Voter model (see plot \textbf{c} of Figure~\ref{fig:magnetization}).

\begin{figure*}[!ht]
\centering
\includegraphics[width=0.75\textwidth]{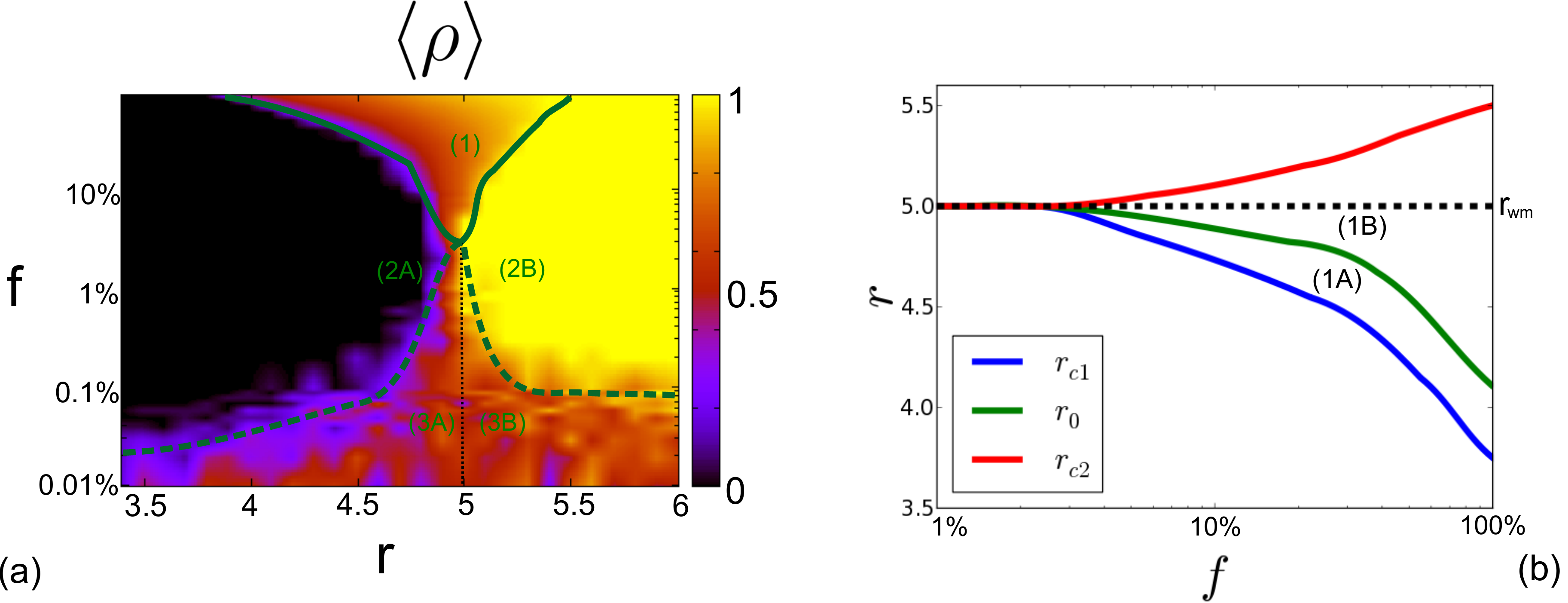}
\caption{\small (Color online) \textbf{a)} Phase diagram, of the average density of cooperators $\langle \rho \rangle$ as a function of the synergy factor $r$ and the fraction of rational agents in the population $f$, i.e., those provided with $K^1 = 0.5$.\textbf{b} Critical thresholds $r_{c1}$, $r_{c2}$ of synergy factors, and the value $r_0$ for which cooperators and defectors coexist in equal number as function of the fraction $f$ of rational agents. Results are averaged over $100$ simulation runs. \label{fig:transition}}
\end{figure*}

\textbf{Heterogeneous Populations. } We now consider the case of a heterogenous population, where a density of agents $f$ ($0 \le f \le 1$) with $K^1=0.5$ is inserted in a population of irrational individuals, performing coin flip to decide their strategy. In this configuration, the level of noise is controlled by the variable $f$, and the lower its value the higher the stochasticity in the population. As shown in Figure~\ref{fig:transition}, the phase diagram obtained as a function of the different values of noise is qualitatively comparable to the previously considered case. 
As $f$ goes to $1$ the PGG turns its behavior to the expected one for a population composed of only rational individuals (i.e., $r_{c1} = 3.75$ and $r_{c2} = 5.5)$.
Remarkably, the outcomes shown in Figure~\ref{fig:transition} suggest that for values as small as $f \sim 3\%$, the PGG shows an active phase (i.e., the network reciprocity still holds). Thus, very few rational agents are able to provide the population an overall rational behavior at equilibrium. See plot \textbf{b} of Figure~\ref{fig:transition} to observe the scaling for the critical values of the synergy factor: $r_{c1}$ (at which cooperators first appear), $r_{c2}$ (at which defectors disappears), and $r_0$ (where cooperators and defectors coexist in equal proportion).

\section{Discussion and Conclusion}\label{sec:conclusions}
The aim of this work is to provide a detailed study of the role of noise in the PGG by the lens of statistical physics. Notably, the proposed model allows to define a clear relation between the noise introduced at the microscopic level in the process named strategy revision phase, and the macroscopic behavior of a population. To achieve this goal, we start from the theoretical considerations presented in~\cite{szabo01,szabo02}, then considering a richer scenario and controlling the noise in two different cases: a homogeneous population (i.e. all agents have the same degree of rationality) and a heterogeneous one (i.e. more degrees of rationality are considered).
The phase diagram resulting from numerical simulations shows the influence of the synergy factor $r$ and of the noise in the macroscopic behavior of the population. So, beyond confirming results reported in works as~\cite{szabo01,moreno02}, our investigation extends to further insights.
Notably, the phase diagram (see Figure~\ref{fig:figure_1}) shows many interesting regions. For a finite range of values of low noise, there exists a second order phase transition between two absorbing states as a function of $r$, with the presence of a metastable regime between them (region $(1)$ ---see plot \textbf{a} of Figure~\ref{fig:figure_1}). 
For higher values of noise the active phase and network reciprocity disappears (regions $(2a)$ and $(2b)$  ---see plot \textbf{a} of Figure~\ref{fig:figure_1}) and the system always reaches an ordered state. In particular, cooperation (defection) is usually reached if $r$ is greater (smaller) than the group size $G=5$, even if fluctuations are possible next to the critical point due to the finite size of the system. As the level of noise increases, the system approaches the behavior of a classical voter model (region $(3)$ ---see plot \textbf{a} of Figure~\ref{fig:figure_1}), where either one of the two ordered phase is reached no matter the value of the synergy factor.
From the analysis of the heterogenous population case, we note that even a very small density $f$ of rational agents, $f \approx 3\%$, allows to observe a network reciprocity effect. In such sense, beyond the physical interpretation of our results, we deem important to highlight that, from the perspective of EGT and from that of sociophysics, the PGG is a system that 'correctly works' even in the presence of few rational players. Here, saying that the system 'correctly works' means that the equilibrium predicted for given a $r$ by the analysis of the Nash equilibria of the system in the well-mixed approximation is achieved.
Finally, it might be interesting to investigate the behavior of the PGG for populations with greater heterogeneity in the rationality of the agents, and possibly change it over time. As a possible interesting case, we suggest that where agents modify their $K$ due to a contact process, in a thermalization-like process. Even if the average value of $K$ in the population is constant, we reckon that differences might emerge with regard to the final steady state of the corresponding homogeneous population, due to the existence of a different initial transient where the distribution of $K$ is non-trivial.
To conclude, our work extends results reported in previous analyses and, moreover, aims to define a statistical physics interpretation of the spatial Public Goods Game. In particular as shown in~\cite{javarone01} an analytical description, defined as the population were a classical spin system, may shed new light and even explain the behavior of those EGT models studied only by a computational approach.
\section*{Acknowledgments}
The authors wish to thank Matjaz Perc for his helpful comments and suggestions.

\end{document}